\documentclass[11pt,a4paper]{article}

\usepackage{tikz}
\usepackage{tikz-feynman}
\tikzfeynmanset{compat=1.1.0}

\usepackage{cite}
\usepackage{amsmath, amsthm, amssymb,slashed}

\usepackage{yfonts}

\usepackage{subcaption}
\usepackage{xfrac}  
\usepackage{fancyhdr}
\usepackage{mathbbol}
\usepackage{url}
\usepackage{color}
\usepackage{bbm}
\usepackage{rsfso}
\usepackage{dsfont}
\usepackage{bm}
\usepackage{amsfonts}
\usepackage{mathrsfs}
\usepackage{slashed}
\usepackage{wasysym}
\usepackage{ytableau}
\usepackage[hidelinks]{hyperref}
\usepackage{cancel}
\usepackage[normalem]{ulem}

\usepackage{graphicx}

\def\bl{\begin{equation}\begin{aligned}}
\def\el{\end{aligned}\end{equation}}
\def\beal{\begin{align}}
\def\eal{\end{align}}
\def\be{\begin{equation}}
\def\ee{\end{equation}}
\def\bpm{\begin{pmatrix}}
\def\epm{\end{pmatrix}}
\def\bsm{\begin{bmatrix}}
\def\esm{\end{bmatrix}}
\def\bvm{\begin{vmatrix}}
\def\evm{\end{vmatrix}}
\def\bVM{\begin{Vmatrix}}
\def\eVM{\end{Vmatrix}}
\def\bea{\begin{eqnarray}}
\def\eea{\end{eqnarray}}

\def\Tr{{\rm Tr}}
\def\1{{\bf 1}}
\def\2{{\bf 2}}
\def\3{{\bf 3}}
\def\4{{\bf 4}}

\usepackage{indentfirst}

\newcommand{\OMIT}[1]{}

\numberwithin{equation}{section}

\usepackage{mathtools}

\usepackage{fontawesome}
\usepackage{microtype,shapepar,xcolor}
\title{Light scalar mesons in $D^+,D_s^+$ decays into three pseudoscalars}
\date{\today}
\author{L. Maiani, A.D. Polosa and V. Riquer
\\ {\it $^*$Sapienza University of Rome and INFN, Piazzale Aldo Moro 2, I-00185, Italy}   }
\begin{document}
\maketitle

\begin{abstract}
In response to recent experimental findings from the BES III and LHCb collaborations concerning the decay of open charm mesons into three pseudoscalars, we present evidence suggesting that the lightest scalar mesons can be characterized as diquark-antidiquark states mixed with the next-to-light, $q\bar q$, scalar mesons through six-fermion, instanton-induced interactions.  
\end{abstract}

\section{Introduction}

Recently, BES III~\cite{BESIII:2020ctr,BESIII:2022vaf} and LHCb~\cite{LHCb:2022pjv,LHCb:2022lja}  have provided detailed data on $D_s^+$ and $D^+$, $S$-wave decays into three pseudoscalar mesons
\begin{align}
&(a)~D_s^+ \to K^+ K^- \pi^+ \quad  \quad 
(b)~D_s^+ \to \pi^+ \pi^- \pi^+ 
\quad \text{(Cabibbo allowed)} \label{allowed} \\
&(c)~D^+ \to \pi^+ \pi^- \pi^+ \quad   \,\,\,\,\;\;\;\,
(d)~D_s^+ \to K^+ \pi^- \pi^+ 
\quad \text{(Cabibbo suppressed)} \label{suppressed}
\end{align}
interpreted as decays into one pseudoscalar  and one scalar meson, followed by the decay of the scalar into two pseudoscalar mesons. 

The final states of these decays have been analyzed in terms of superpositions of pseudoscalar mesons and resonances, encompassing all meson resonances that undergo two-body decays. The estimated fraction of two mesons in the $S$-wave final state for~\eqref{allowed}(a) is approximately 24\%, while for~\eqref{allowed}(b) and~\eqref{suppressed}(c), it is 85\% and 62\%, respectively. For~\eqref{suppressed}(d), the fraction is 46\%.

Data provide valuable insights into the structure of the scalar mesons and may contribute to the resolution of the enigma surrounding the internal structure of the lightest scalar nonet.

In this paper, we employ the experimental data on decays~\eqref{allowed} and~\eqref{suppressed} to assess the tetraquark hypothesis for the composition of the light scalar mesons $\sigma, f_0,$ and $\kappa$. Furthermore, we investigate their  mixing to the nonet of $q\bar q$ scalar mesons proposed in~\cite{Jaffe:1976ih,Maiani:2004uc,tHooft:2008rus,Fariborz:2008bd}.

In Sect.~\ref{scalmes} we review the arguments for describing the scalar meson multiplets as: $i)$ a heavier, $P$-wave $q\bar q$, nonet and $ii)$ a light diquark-antidiquark nonet,  the two being mixed by interactions generated by QCD instantons~\cite{tHooft:1999cta,Shifman:1979uw}. 

In Sect.~\ref{C_allowedK}, we consider the quark diagrams describing Cabibbo allowed decay, Eq.~\eqref{allowed}$(a)$. We  introduce four phenomenological amplitudes, estimating three of them from the observed branching ratios of the $D^+_s \to K^+ K^- \pi^+$ decay. 
In Sect.~\ref{C_allowedPi} we give an upper bound to the missing fourth amplitude from the non-observation of $\sigma$ production in $D^+_s \to \pi^+\pi^-\pi^+$decay, Eq.~\eqref{allowed}(b), and predict the branching ratios of the other decay modes.  In Sects.~\ref{C_suppressedPi} and \ref{C_suppressedK},  we extend our considerations to the Cabibbo suppressed decays, Eq.~\eqref{suppressed}.  In this scheme we are able to  reproduce  the relative branching fractions of \eqref{allowed}$(b)$ and \eqref{suppressed}$(c)$ decays. 

For the last decay, \eqref{suppressed}$(d)$,  the comparison of our predictions  with the data from \cite{BESIII:2022vaf} is not so good. In part this may be due to  rather large experimental errors. Also, unlike previous cases, BESIII fits the low energy $K^+\pi^-$ data by a continuum distribution, rather then the Breit-Wigner corresponding to the $\kappa(700)$ resonance, as would be required by our model.

A more conclusive statement will require larger statistics and more precise data, fitted consistently with the two nonets of scalar resonances assumed in the present analysis.

\section{Scalar Mesons\label{scalmes}}
{\bf{\emph{ The higher  $q\bar q$ nonet.}}} Positive-parity $q \bar q$ scalar mesons are expected to occur in  $P$-wave~\cite{Borchi:1965sdr} and therefore be heavy. 
$P$-wave scalar mesons have indeed been observed above 1 GeV, see Tab.~\ref{heavscal}, together with the other states required by the composition of orbital and spin angular momenta~\cite{ParticleDataGroup:2024cfk}.
\begin{align}
& J=s_q+s_{\bar q}+L,~L=1, \,\,\, J^{PC}=0^{++}\\
& (a_0(1450),\dots),~1^{++}~(a_1(1260),\dots),~1^{+-}~(b_1(1235),\dots),~2^{++}~(a_2(1320),\dots). \notag
\end{align}

Scalar, $q \bar q$, mesons can be classified in the SU(3)$_F$ representations ${\bf {8}}\oplus {\bf{1}}$.
~In particular, $f_0(1370)$ and $f_0(1710)$ are assumed to be associated to the $\bar u u + d \bar d$ and $s \bar s$ configurations, respectively.  The additional singlet in Tab.~\ref{heavscal}, $f_0(1500)$, has been proposed to be predominantly a glueball~\cite{Amsler:1995td,Shen:2002nv,Close:2005vf,Close:2002zu} and will not be considered in the following.

\begin{table}[ht]
\centering
    \begin{tabular}{|c|c|c|c|}
   \hline 
{\footnotesize PDG name} &{\footnotesize M}& {\footnotesize $\Gamma$} & {\footnotesize Partial Rates}\\
 \hline
\hline
  {\footnotesize $f_0(1370)$ }& {\footnotesize $1200-1500$}  & {\footnotesize $50-500$}&{\footnotesize $2 \pi$ dominant} \\
  \hline
{\footnotesize $a_0(1450)$}  &{\footnotesize $1474\pm 19$}  & {\footnotesize $265\pm 13$} & \\
  \hline
  {\footnotesize  $K_0(1430)$} &{\footnotesize $1425\pm 50$}  & {\footnotesize $270\pm 80$} &    \\    
   \hline
   {\footnotesize $f_0(1500)$} & {\footnotesize $1522\pm 25$}  & {\footnotesize $108\pm 33$} & \footnotesize{glueball?}\\
   \hline
   {\footnotesize $f_0(1710)$} &{\footnotesize $1733\pm 8$}  & {\footnotesize $150\pm 11$} & {\footnotesize{$K\bar K$ dominant}}\\
\hline 
\end{tabular}
 \caption {\footnotesize {
 Mass and widths of the  heavier scalar mesons from PDG 2024~\cite{ParticleDataGroup:2024cfk}.}}
 \label{heavscal}
 \end{table}

The $q\bar q$ nonet is represented by the matrix
\be
S^\prime(q\bar q) 
=
\bpm
\frac{f^{\prime }+a^{\prime 0}}{\sqrt{2}} 
  & a^{\prime +} &  \kappa^{\prime +}  \\ 
a^{\prime -} & \frac{f^{\prime}-a^{\prime 0}}{\sqrt{2}}  
  & {\kappa}^{\prime 0} \\ 
\kappa^{\prime -} & \bar{\kappa}^{\prime 0} &  f^{\prime\prime}
\epm
\label{s2q}
\ee
with the shorthand notation  $f^{\prime }=f_0(1370),~f^{\prime\prime}=f_0(1710)$; when $f_0$ is found, we mean $f_0=f_0(980)$, the lightest of $f_0,f^\prime, f^{\prime\prime}$.

{\bf{\emph{The lower  tetraquark nonet.}}} Following Jaffe~\cite{Jaffe:1976ih} and successive developments~\cite{Maiani:2004uc,tHooft:2008rus,Fariborz:2008bd}, we assume that the lightest 
scalar multiplet is made by spin zero diquark-antidiquark  bound states forming the SU(3)$_F$ nonet. 
We will define the spin-0 diquark as\begin{equation}
[qq]_{i\alpha}= \epsilon_{ijk}\epsilon_{\alpha \beta\gamma} (\bar q_c)^{j\beta}\gamma_5 q^{k\gamma}.
\label{defdq}
\end{equation}
where latin indices are for flavor, greek for color and the subscript $c$ is for charge-conjugation.\footnote{where $\bar q_c\gamma_5 q=-iq^T(1\otimes\sigma_2)q$. }  This ensures Fermi statistics: the state is anti-symmetric in color, spin and flavor. The spin-0, $\bm 3_c$ diquark with flavors $a$ and $b$ is  $[q^a q^b]_\alpha$.
The connection between tetraquark states and SU(3)$_F$ matrices is given by the relations
\be
S_j^i=\epsilon_{iab}\epsilon^{jcd}M^{ab}_{cd}
\quad \text{where}\quad M^{ab}_{cd}=[ q^a q^b]_\alpha[\bar q_c \bar q_d]^\alpha
\ee
For example, $S_1^1$ is the diquark combination $[ds][\bar d\bar s]=(f_0-a_0)/\sqrt{2}$, $S_2^2=[us][\bar u\bar s]=(f_0+a_0)/\sqrt{2}$ and $S_3^3=[ud][\bar u\bar d]$ with
\be
S([qq][\bar q \bar q])=
\bpm 
\frac{f_0-a^0 }{\sqrt{2}} & a^+ & \kappa^+\\ a^- &\frac{f_0+a^0 }{\sqrt{2}}& \kappa^0 
\\ \kappa^- &  \bar \kappa^0 & \sigma  
\epm 
\label{tetmatr}
\ee
The unmixed diagonal states are
\begin{align}
&\sigma =[ud][\bar u \bar d]\\
&f_0,a_0=\tfrac{1}{\sqrt{2}}\Big([us][\bar u \bar s]\pm[ds][\bar d \bar s]\Big)
\end{align}

When the mixing angle between the components $(S_1^1+S_2^2)/\sqrt{2}$ and $S_3^3$ is very small we have the structure of an ideally mixed 4-quark nonet. If  we had a mixing close to $90^\circ$  the scalars would be better described  in terms  $q\bar q$ mesons: exchange $f_0\leftrightarrow \sigma$ in~\eqref{tetmatr}. The latter case however is questionable because  of the degeneracy $f_0(s\bar s)\sim a_0(q\bar q)$, in addition to the fact that scalar mesons of the $q\bar q$ kind should be $P$-wave states.

Associated with the chiral symmetry breaking SU(3)$_L\times$SU(3)$_R\to$SU(3)$_V$ we have an octet of pseudoscalar mesons $\Phi$ (pseudo-Goldstone bosons)~\footnote{Where $$\eta_q=\frac{\bar u \gamma_5 u + \bar d \gamma_5 d}{\sqrt{2}} \quad \text{and} \quad \eta_s= \bar s\gamma_5 s$$}
\bea
\Phi = \bpm
\frac{\eta_q+\pi^0}{\sqrt{2}}
& \pi^+ &  K^+ \\ 
\pi^- & \frac{\eta_q-\pi^0}{\sqrt{2}} & K^0 \\ 
 K^- & {\bar K}^0 & \eta_s 
\epm
\label{eq:Phi}
\eea
Scalar mesons  interact with pseudoscalars with  derivative interactions including $\partial_\mu\Phi\equiv\Phi_\mu$. The terms in the interaction Lagrangian involving the scalar and pseudoscalar mesons are 
\be
{\cal L}=A\,\Tr(S\Phi_\mu \Phi^\mu)+B\,\Tr(S)\Tr(\Phi_\mu \Phi^\mu)+
C\,\Tr(S\Phi_\mu)\Tr(\Phi^\mu)+D\,\Tr(S)\Tr(\Phi_\mu )\Tr(\Phi^\mu)
\label{effL}
\ee
In general the parameters $A,B$ can be determined by the analysis of $\pi\pi$ or $\pi K$ scattering whereas $C,D$ can be determined in processes involving $\eta$ and $\eta^\prime$ particles because of the explicit form of $\Tr(\Phi_\mu)$. In the tetraquark picture we consider, for example, a transition like $[ds][\bar d\bar s]\to (s\bar d)(d\bar s)$, where the $d$ and $\bar d$ quarks are being swapped, breaking the diquarks.  Such a transition corresponds, in the notations given above, to $S_1^1\to (\Phi_\mu)^2_3(\Phi^\mu)^3_2$. The interaction term $S_1^1 (\Phi_\mu)^2_3(\Phi^\mu)^3_2$ comes with $2B$ in the Lagrangian ~\eqref{effL}. With $2B$  we have also  the  $S_1^1 (\Phi_\mu)^1_3(\Phi^\mu)^3_1$ interaction, which  corresponds to the subleading transition $[ds][\bar d\bar s]\to (s\bar u)(u\bar s)$ where the $d\bar d$ pair converts into a $u\bar u$ pair. In the following we will include only the leading quark-swap amplitudes. In order to do this, we note that $S_1^1 (\Phi_\mu)^1_3(\Phi^\mu)^3_1$ comes also from the term weighted by  $A$ in~\eqref{effL} and  the combination $2B+A=0$, which is satisfied for $B=1/2$ and $A=-1$, removes this term.  So the leading quark swapping interaction in the tetraquark model (subscript-$f$: flavor-swapping) corresponds to 
\be
{\cal L }_{s}=C_f \Big(\tfrac{1}{2}\,\Tr(S)\Tr(\Phi_\mu \Phi^\mu)-\,\Tr(S\Phi_\mu \Phi^\mu)\Big)=C_f O_f
\ee
Alternatively we can consider transitions in which $[ds][\bar d\bar s]\to (d\bar d)(s\bar s)$, i.e. involving the pseudoscalar singlet. This corresponds to $S_1^1\to (\Phi_\mu)_2^2(\Phi_\mu)_3^3$ which  appears with coefficient $2D$ in~\eqref{effL}. The operator weighted by $D$ involves also terms like  $S_1^1\to (\Phi_\mu)_1^1(\Phi_\mu)_3^3$ which are cancelled by those deriving from the operator weighted by $C$,  if we require $C=-1$ and $D=1/2$. 
The transitions $[ds][\bar d\bar s]\to (s\bar d)(d\bar s)$ and $[ds][\bar d\bar s]\to (d\bar d)(s\bar s)$ represent two alternative quark swaps. In the first case, the $d$ quarks are swapped with the $s\bar s$ pair being spectators, while in the second case, a $d$ and $\bar s$ (or vice versa) are swapped. Under the assumption that there is no distinction between these two quark swappings, the corresponding operators can be grouped with the same effective coefficient
\be
{\cal L }=C_f O_f=C_f\Big( -\Tr(S\Phi_\mu \Phi^\mu)+\tfrac{1}{2}\,\Tr(S)\Tr(\Phi_\mu \Phi^\mu)
-\Tr(S\Phi_\mu)\Tr(\Phi^\mu)+\tfrac{1}{2}\,\Tr(S)\Tr(\Phi_\mu )\Tr(\Phi^\mu)\Big)
\label{effL2}
\ee
where now $O_f$ includes all the four terms in~\eqref{effL}, with only one phenomenological parameter $C_f$ in place of $A,B,C,D$. 

There is an additional interaction of the $\Tr(S\Phi_\mu\Phi^\mu)$ type to be considered though. It is induced by QCD instantons  producing an effective interaction reducing 
the ${\rm U}(N_f)_L \times {\rm U}(N_f)_R$ global symmetry of the quark model 
in the chiral limit to ${\rm SU}(N_f)_L \times {\rm SU}(N_f)_R$ times baryon number.
The effect can be described by the 't Hooft 6-fermion interaction~\cite{tHooft:1999cta,Shifman:1979uw}
\be
{\cal{L}}_I \propto  \det [(q_L)_{i\alpha}(q_R)^j_\alpha]
\ee
where $i$ and $j$ denote flavor indices. With three light quark flavors, ${\cal L}_I$ is proportional to the product of three 
quark and three antiquark fields, anti-symmetrized in flavor and color, and it includes 
a term of the type
\be
\text{Tr}(J^{[4q]} J^{[2q]} )~ \label{instact}
\ee
where
\be
(J^{[4q]})^i_{j} =  [{\bar q} {\bar q}]^{i\alpha}  [ q q]_{j\alpha}\qquad  (J^{[2q]})^i_j =  {\bar q}_{j\alpha} q^{i\alpha}
\ee
The two components of the effective Lagrangian~\eqref{instact} can annihilate a tetraquark and create a $q\bar q$ and viceversa. Therefore the interaction Lagrangian~\eqref{effL2} can be extended to include the instanton induced interaction as in 
\be
{\cal L}=C_f O_f+C_I  O_I 
\label{effL3}
\ee
The second term induces transitions $4q\leftrightarrow 2q$; notice that   the $O_I$ operator is itself of the form $\Tr(S\Phi_\mu\Phi^\mu)$.
This follows from the chiral realization of the currents in~\eqref{instact} and 
\be
\langle PP|\text{Tr}(J^{[4q]} J^{[2q]} )|S\rangle \propto \Tr(S\Phi_\mu\Phi^\mu)+\dots
\ee
where, as commented in~\cite{tHooft:2008rus} (Eq. (16)), the dots denote higher-order terms in $\Phi$ and in the chiral expansion.  At low energy, instanton effects are expected to be sizeable.

The instanton interaction induces the mixing of the two nonet fields of the form
\be
{\cal L}_{S^\prime S}=\gamma \, \text{Tr}(S^\prime S)  \label{mix}
\ee
where $S^\prime$ is the conventional scalar $q\bar q$ nonet. 

The  mixing provided by the effective QCD action, Eq.~\eqref{mix}, among other effects, is responsible for the otherwise forbidden decay $f_0(980) \to \pi^+\pi^-$ of the $f_0$ tetraquark~\cite{tHooft:2008rus}.  Indeed the Zweig rule allowed decays are $f_0\to K\bar K $ and $f^\prime \to \pi\pi$, but not $f_0\to \pi\pi$. 

Mixing can go either way ($\lambda<1$)
\begin{align}
& f_0(980)\equiv f_0= f_{0,4q}+\lambda f^\prime_{2q}      \label{instmix1}   \\
& f_0(1370)\equiv f^\prime =  -\lambda f_{0,4q}+ f^\prime_{2q} \label{instmix2}
\end{align}
where we mean that the mass eigenstate $f_0$ is essentially the 4-quark eigenstate $f_{0,4q}$ mixed with a 2-quark component $f^{\prime}_{2q}$ (the prime makes reference to $S^\prime$). Similarly the mass eigenstate $f^\prime$ is essentially $f^\prime_{2q}$. Thanks to the mixing, it acquires a tetraquark component and it displays the corresponding decay modes, e.g. $f^{\prime\prime}\to \pi^+\pi^-$ via its mixing with $\sigma$, and can be created by 4 quark operators, as we shall discuss in Sect.~\ref{C_allowedK}.

The ordering $M[a_0(1450)]>M[K_0(1430)]$ can also be attributed to the mixing with the corresponding light scalar mesons $a_0(980)$ and $\kappa(700)$ induced by the lagrangian Eq.~\eqref{mix}. This analysis provides the estimate (see~\cite{Black:1999yz},~\cite{Ali2019})
\be
\gamma \simeq 0.7 ~\text{GeV}^2 \label{numix}
\ee

\section{Cabibbo allowed $S$-wave decay: $D^+_s~\to K^+ K^-\pi^+$\label{C_allowedK}}

\begin{figure}[h!]
\centering
\scalebox{0.8}{
\begin{tikzpicture}
  \begin{scope}[shift={(-4,0)}]
    \shade[ball color=gray!50] (0,0) circle (0.3);
    \node at (-0.7,0) {$D_s^+$};
      \node at (2,-1.7) {$(b)$};

    \begin{feynman}
      \vertex (a) at (0,0);
      \vertex (f1) at (3,1.2);
      \vertex (f2) at (3,-1.2);

      \vertex (b) at (1.5,0);
      \vertex (f3) at (3,0.6);
      \vertex (f4) at (3,-0.6);

      \vertex (mid) at ($(1,1.2)!0.5!(f1)$);
      \vertex (p) at ($(mid)+(1.5,1)$);
      \vertex (f5) at (5,2.8);
      \vertex (f6) at (5,1.6);

      \node at ($(f1)+(0.3,0)$) {$s$};
      \node at ($(f2)+(0.3,0)$) {$\bar s$};
      
      \node at ($(f3)+(0.3,0)$) {$u_2$};
      \node at ($(f4)+(0.3,0)$) {$\bar u$};
     
      \node at ($(f5)+(0.3,0)$) {$u_1$};
      \node at ($(f6)+(0.3,0)$) {$\bar d$};

  \node at ($(p)+(2.5,0)$) {$\pi^+$};

      \diagram* {
        (a) -- [fermion, out=90, in=180, thick] (f1),
        (a) -- [anti fermion, out=-90, in=180, thick] (f2),

        (b) -- [fermion, out=90, in=180, thick] (f3),
        (b) -- [anti fermion, out=-90, in=180, thick] (f4),

        (mid) -- [photon, thick] (p),
        (p) -- [fermion, out=90, in=180, thick] (f5),
        (p) -- [anti fermion, out=-90, in=180, thick] (f6),
      };
    \end{feynman}
  \end{scope}


  \begin{scope}[shift={(4,0)}]
    \shade[ball color=gray!50] (0,0) circle (0.3);
    \node at (-0.7,0) {$D_s^+$};
       \node at (2,-1.7) {$(c)$};

    \begin{feynman}
      \vertex (a) at (0,0);
      \vertex (f1) at (3,1.2);
      \vertex (f2) at (3,-1.2);

      \vertex (b) at (1.5,0);
      \vertex (f3) at (3,0.6);
      \vertex (f4) at (3,-0.6);

      \vertex (mid) at ($(1,1.2)!0.5!(f1)$);
      \vertex (p) at ($(mid)+(1.5,1)$);
      \vertex (f5) at (5,2.8);
      \vertex (f6) at (5,1.6);
      
       \node at ($(f1)+(0.3,0)$) {$s$};
      \node at ($(f2)+(0.3,0)$) {$\bar s$};
      
      \node at ($(f3)+(0.3,0)$) {$d$};
      \node at ($(f4)+(0.3,0)$) {$\bar d_2$};
     
      \node at ($(f5)+(0.3,0)$) {$u$};
      \node at ($(f6)+(0.3,0)$) {$\bar d_1$};
  \node at ($(p)+(2.5,0)$) {$\pi^+$};

      \diagram* {
        (a) -- [fermion, out=90, in=180, thick] (f1),
        (a) -- [anti fermion, out=-90, in=180, thick] (f2),

        (b) -- [fermion, out=90, in=180, thick] (f3),
        (b) -- [anti fermion, out=-90, in=180, thick] (f4),

        (mid) -- [photon, thick] (p),
        (p) -- [fermion, out=90, in=180, thick] (f5),
        (p) -- [anti fermion, out=-90, in=180, thick] (f6),
      };
    \end{feynman}
  \end{scope}
\end{tikzpicture}
}
\caption{{\footnotesize{$D_s$ decay, Cabibbo allowed,  with a quark pair from vacuum.}} \label{f2}}
\end{figure}
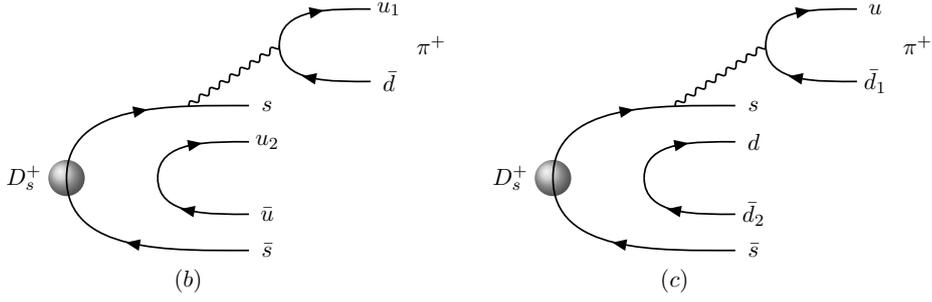

The relevant quark Feynman diagrams for Cabibbo allowed $D^+_s$ decays are reported in Fig.~\ref{f1} and in Fig.~\ref{f2}, when a quark pair is created from vacuum. 
Considering the decay $D^+_s\to K^+ K^- \pi^+$, we observe that

\begin{figure}[h!]
\centering
\scalebox{0.8}{ 
\begin{tikzpicture}
  \shade[ball color=gray!50] (0,0) circle (0.3);
  \node at (-0.7,0) {$D_s^+$};
    \node at (2,-1.7) {$(a)$};
  \begin{feynman}
    \vertex (a) at (0,0);
    \vertex (f1) at (3,1.2);
    \vertex (f2) at (3,-1.2);
    \vertex (f5) at (5,2.8);
    \vertex (f6) at (5,1.6);
    
     \node at ($(f1)+(0.3,0)$) {$s$};
      \node at ($(f2)+(0.3,0)$) {$\bar s$};
      
       \node at ($(f5)+(0.3,0)$) {$u$};
      \node at ($(f6)+(0.3,0)$) {$\bar d$};
 
      \node at ($(a)+(3,0)$) {$J^P=0^+,\; L=1$};

    \vertex (mid) at ($(1,1.2)!0.5!(f1)$);
    
    \vertex (p) at ($(mid)+(1.5,1)$);

     \node at ($(p)+(2.5,0)$) {$\pi^+$};

    \diagram* {
      (a) -- [fermion, out=90, in=180, thick] (f1),
      (a) -- [anti fermion, out=-90, in=180, thick ] (f2),
    
      (mid) -- [photon, thick] (p),
      
      (p) -- [fermion, out=90, in=180, thick] (f5),
      (p) -- [anti fermion, out=-90, in=180, thick ] (f6)
        
    };
  \end{feynman}
\end{tikzpicture}
}  
\caption{\footnotesize{$D_s$ decay,  Cabibbo allowed, tree-level diagram.}\label{f1}}
\end{figure}
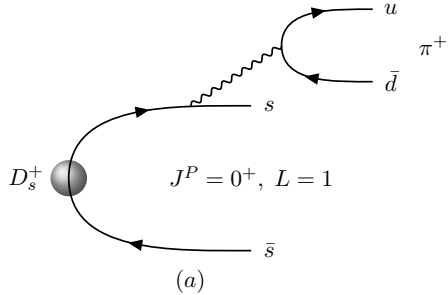

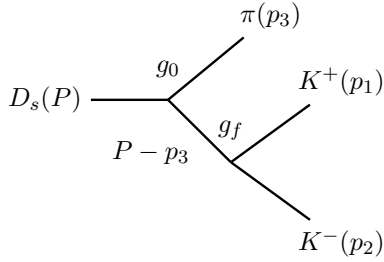
\begin{figure}[h!]
\centering
\scalebox{0.9}{
\begin{tikzpicture}
  \begin{feynman}
    \vertex (a1) {\(D_s(P)\)};
      \vertex [right=1.8cm of a1] (b1);
    \vertex [above =0.2cm of b1]  {\(g_0\)};  
    \vertex [above right=1.3cm of b1] (f1a) {\(\pi(p_3)\)};
    \vertex [below right=1.3cm of b1] (c1);
    \vertex [above right=1.2cm of c1] (f2a) {\(K^+(p_1)\)};
    \vertex [below right=1.2cm of c1] (f3a) {\(K^-(p_2)\)};
     \vertex [above =0.2cm of c1] {\(g_f\)};

   \diagram* {
      (a1) -- [plain, line width=1pt] (b1) -- [plain, line width=1pt] (f1a),
      (b1) -- [plain, line width=1pt, edge label'=\(P-p_3\)] (c1),
      (c1) -- [plain, line width=1pt] (f2a),
      (c1) -- [plain, line width=1pt] (f3a),
    };

      \end{feynman}
\end{tikzpicture}
}
\caption{{\footnotesize{Feynman diagram representation of the $D_s\to (K\bar  K)_f\pi$ decay amplitude}}. \label{f3}}
\end{figure}

 \begin{enumerate} 
 \item Diagram $(a)$ may produce the $f^{\prime\prime}$ meson via its $q\bar q$ component (amplitude $A_f$) and the $\sigma$ meson via its instanton mixing to the $s\bar s$ meson (amplitude $A_I$).   The amplitude $A_I$ and $A_f$ comprise the weak interaction pion production. The amplitudes $A_f$ and $A_I$ are generated by interactions of the $C_f\, O_f$ and $C_I\, O_I$ type, as in~\eqref{effL3}.  However, $A_I$ cannot be determined from the rates $\Gamma(D_s\to K\bar K\pi)$ since $\sigma$  cannot decay into $K\bar K$.
  
\item Diagrams $(b)$ and $(c)$ produce $f_0$ via its dominant tetraquark component (amplitude $B_f$) and $f^{\prime}$ via its mixing generated tetraquark component, with subsequent decays in $K^+K^-$ (amplitude $B_I$). The amplitudes $B$ comprise the weak interaction pion production. The amplitudes $B_f$ and $B_I$ are generated by interactions of the $C_f\, O_f$ and $C_I\, O_I$ type, as in~\eqref{effL3}.
\end{enumerate}

 The amplitudes $A_f,B_f,B_I$ are computed by comparison to data as follows. We will evaluate the Feynman diagram in Fig.~\ref{f3} representing the $D_s\to f\pi\to K\bar K\pi$ decay. Its square amplitude  is given by
\be
|M|^2=\frac{g_0^2 \, g_f^2}{\left(m_f\Gamma_f\right)^2+\left(m_f^2-m_{12}^2\right)^2}
\label{emm2}
\ee 
and the rate $\Gamma(D_s\to f\pi\to K\bar K\pi)$ is computed applying the standard 3-body phase space formula
\be
d\Gamma=\frac{1}{(2\pi)^3}\frac{1}{32M^3}\, |M|^2\, dm_{12}^2\, dm_{23}^2 
\label{3body}
\ee
with the $m_{12}^2$ and $m_{23}^2$ variables to be integrated over the-Dalitz plot surface.  The parameters $m_f$ and $\Gamma_f$ are the mass and the total width of the intermediate scalar resonance. The coupling $g_f$ is determined by the partial width of the scalar meson to decay into $K^+K^-$, when this channel is available
\be
 g_f=\sqrt{\frac{ 8\pi m_f^2\, {\cal B}(f\to K\bar K)\,  \Gamma_f}{p^\star(m_f,m_K,m_K)}}
\ee
where $p^\star$ is the decay momentum of the particle of mass $M$ into $m_1$ and $m_2$,   $p^\star(M,m_1,m_2)=\sqrt{\lambda(M^2,m_1^2,m_2^2)}/2M$, $\lambda$ being the K\"all\'en triangular function. The coupling $g_0$ contains the aforementioned $A_{f,I},B_{f,I}$ amplitudes. We have access to 
\bea
 &g_0=(B_f\sqrt{2})\sqrt{m_{D_s}}& \quad \text{for}\quad D_s\to f_0 \pi \to K\bar K\pi\label{44}\\
 &g_0=(B_I\sqrt{2})\sqrt{m_{D_s}}&\quad \text{for}\quad D_s\to f^\prime\pi\to K\bar  K\pi\label{45}\\
 &g_0=A_f\sqrt{ m_{D_s}} &\quad \text{for}\quad D_s \to f^{\prime\prime}\pi \to K\bar  K\pi
\eea
see the notation for scalar mesons introduced after~\eqref{s2q}. The introduction of the $\sqrt{m_{D_s}}$ dependency allows  $\Gamma, A,B$ to be {\bf{\it{pure numbers}}}. The factors $\sqrt{2}$ on the first two equations arise because the sum of diagrams $(b)+(c)$ corresponds to $\sqrt{2}f_0$, for $B_f$ and $\sqrt{2}f^\prime$, for $B_I$, see Appendix. 
The physical parameters used in the previous calculation are extracted from  Tab.~\ref{tab:res2} and Tab.~\ref{tab:resnext}.
\begin{table}[ht]
\begin{center}
\begin{tabular}{|l||c|c|c|c|cl||}
\hline
{\small Processes  }& {\small  $m_f$ (MeV)} &  {\small $\Gamma_f$ (MeV) }& {\small $p^\star$~(MeV)}&{\small {\cal BR}} 
\\ 
 \hline\hline
{\small $\sigma   \to \pi^+\pi^- $ }& {\small $475$} & {\small $275$} &  {\small  $192$ } &{\small 1$\cdot$~2/3}
\\ \hline
{\small $\kappa^+ \to  K^0 \pi^+ $} &{\small  $845$} &{\small $468$ }&   {\small $253$}&{\small 1$\cdot$2/3} 
\\  \hline
{\small $f_0 \to \pi^+\pi^- $  }  & {\small $990$} & {\small $45$ }&  {\small $475$ }&{\small 0.75$\cdot$~2/3}
\\
{\small $f_0 \to    K^+ K^- $ }   &{\small  $990$ }& {\small $45$} &{\small $36$ }&{\small 0.25/2}
\\\hline
{\small $a_0 \to  \pi^0 \eta$  }  & {\small $980$} &  {\small $75$} &{\small $317$}&{\small 0.85} 
\\
{\small $a_0 \to K^+ K^-$    }    &{\small $980$} &  {\small  $75 $ }&{\small $87-33$ }&{\small 0.15/2} 
\\\hline\hline
\end{tabular}  
\caption{\footnotesize{Experimental parameters for the lightest scalar meson decays~\cite{ParticleDataGroup:2024cfk}. In the BR column, the first figure is the PDG branching ratio, which includes all charge states, to be multiplied by the figure next to it to obtain the charge combination indicated in the first column.}}
\label{tab:res2}  
\end{center}
\end{table} 


\begin{table}[ht]
\begin{center}
\begin{tabular}{|l||c|c|c|c|cl||}
\hline
{\small Processes  }& {\small  $M_S$ (MeV)} &  {\small $\Gamma_S$ (MeV) }& {\small $p^\star$~(MeV)} &{\small $BR$} 
\\ 
 \hline\hline
{\small $f_0(1370)   \to \pi^+\pi^- $}~
&{\small $1345$}& {\small $350$}& {\small  $672$ }&{\small 0.26~$\cdot$ 2/3} 
 \\ \hline
{\small $f_0(1370)   \to K^+ K^- $ }& {\small $"$} & {\small $"$} &  {\small  $475$ }&{\small 0.021/2} 
\\ \hline\hline
{\small $a^{ 0}_0(1450) \to  \pi^0 \eta$  }  & {\small $1439$} &  {\small $258$} &{\small $606$} &{\small 0.093} 
\\  \hline
{\small $a^{0}(1450) \to  \pi^0 \eta^\prime(958)$  }  & {\small $"$} &  {\small $"$} &{\small $383$}&{\small 0.033}  
 \\ \hline
{\small $a^{ 0}_0(1450) \to K^+ K^-$    }    & {\small $"$} &  {\small $"$} &{\small $523$}&{\small 0.083/2} 
\\ \hline \hline
{\small $K^{+}_0(1430) \to  K^0 \pi^+ $} &{\small  $1425$} &{\small $270$ }&   {\small $621$}&{\small 0.93~$\cdot$~2/3} 
\\  \hline
{\small $K^{ +}_0(1430) \to  K^+\eta $} &{\small  $"$} &{\small $"$ }&   {\small $490$}&{\small 0.086} 
\\  \hline \hline
{\small $f^{}_0(1710) \to \pi^+\pi^- $  }  & {\small $1733$} & {\small $150$ }&  {\small $855$ }&{\small 0.039 $\cdot$~2/3}
\\ \hline
{\small $f^{}_0(1710) \to    K^+ K^- $ }   &{\small  $"$ }& {\small $"$} &{\small $711$  }&{\small 0.36/2}
\\  \hline \hline
\end{tabular}  
\caption{\footnotesize{Experimental parameters~\cite{ParticleDataGroup:2024cfk} for the next-to-light scalar mesons. 
In the BR column, the first figure is the PDG branching ratio, which includes all charge states, to be multiplied by the figure next to it to obtain the charge combination indicated in the first column. }}
\label{tab:resnext}
\end{center}
\end{table} 


The experimental data to confront with are reported in the following Table~\ref{brfrac1}. 
\begin{table}[ht]
\centering
    \begin{tabular}{|c|c|c|}
     \hline
{\footnotesize {$D_s^+ \to ~ \pi^+K^+K^-$ }}& {\footnotesize{ BR(\%)}} &{\footnotesize{$(100/5.45)\cdot$BR}}   \\
 \hline
\hline
{ \footnotesize { $\pi^+K^+K^-$} }& {\footnotesize{ $5.45$ }}&{\footnotesize {$100$}}\\ \hline
   {\footnotesize {$f_0(980)\pi^+$, $f_0\to K^+K^-$}}&{\small $1.1$}& {\footnotesize {$20.5$}} \\ \hline
    {\footnotesize {$a_0^0(980)\pi^+$, $a_0^0(980)\to K^+K^-$}}& {\footnotesize {$0.055$}}&{\footnotesize {$1.0$}} \\ \hline
     {\footnotesize {$f_0(1370)\pi^+$, $f_0(1370)\to K^+K^-$}}&{\small $7.1~\cdot 10^{-2}$}& {\footnotesize {$1.32$}} \\ \hline
     {\footnotesize {$f_0(1710)\pi^+$, $f_0(1710)\to K^+K^-$}}&{\small $6.7~\cdot~10^{-2}$}& {\footnotesize {$1.24$}} \\ \hline
  \hline
\end{tabular}
 \caption {\footnotesize {Branching fractions of resonant contributions to the decay: $D_s^+\to \pi^+(K^+K^-)_{S}$, from PDG 2024~\cite{ParticleDataGroup:2024cfk}.}}
 \label{brfrac1}
 \end{table}
Using Table~\ref{brfrac1} and equating $\Gamma(D_s\to f_0\pi\to K\bar  K\pi)=20.5$, $\Gamma(D_s\to f^\prime\pi\to K\bar  K\pi)=1.32$ and  $\Gamma(D_s\to f^{\prime\prime}\pi\to K\bar  K\pi)=1.24$  we obtain\footnote{We will compute only ratios of widths in the following.} the three parameters (expressed as pure numbers in the parameterization formulated above) $A_f,B_f,B_I$ which we will be using to study the $D_{(s)}\to 3\pi$ decays
\be
|A_f|=46.1,~|B_f|=55.2, ~|B_I|=96.7 \label{values}
\ee 
showing that instanton interactions in this scheme are of the same size of quark-swapping ones.

\section{Cabibbo allowed $S$-wave decay: $D^+_s~\to \pi^+\pi^+\pi^-$\label{C_allowedPi}}

The quark diagrams are the same as $(a),(b),(c)$ in Figs.~\ref{f1} and~\ref{f2}.  In this case, we observe that 
 \begin{enumerate}
 \item Diagram $(a)$ is related to the production of $f^{\prime\prime}$ proportionally to the amplitude $A_f$.  The instanton interaction allows the $\sigma$ in diagram $(a)$ so that we can determine an upper value for the amplitude $A_I$ by comparison to data on $D_s\to 3\pi$ decays,  which indicate a very stringent upper bound on the $\sigma$ in $D_s\to 3\pi$ decays.  
 \item Diagrams $(b)$ and $(c)$ are related to the production of $f_0$ through amplitude $B_f$.
 The production of $f^\prime$ is proportional to  amplitude $B_I$. 
\end{enumerate}

\begin{figure}[h!]
\centering
\begin{tikzpicture}
  \begin{feynman}
    \vertex (a1) {\(D_s(P)\)};
      \vertex [right=1.8cm of a1] (b1);
    \vertex [above right=1.3cm of b1] (f1a) {\(\pi^+(p_3)\)};
    \vertex [below right=1.3cm of b1] (c1);
    \vertex [above right=1.2cm of c1] (f2a) {\(\pi^+(p_1)\)};
    \vertex [below right=1.2cm of c1] (f3a) {\(\pi^-(p_2)\)};

   \diagram* {
      (a1) -- [plain, line width=1pt] (b1) -- [plain, line width=1pt] (f1a),
      (b1) -- [plain, line width=1pt, edge label'=\(P-p_3\)] (c1),
      (c1) -- [plain, line width=1pt] (f2a),
      (c1) -- [plain, line width=1pt] (f3a),
    };

    \node at ($(b1)!0.5!(b1)+(4.3cm,0)$) {{\huge $+$}};

    \vertex [right=7.5cm of a1] (a2) {\(D_s(P)\)};
    \vertex [right=1.8cm of a2] (b2);
    \vertex [above right=1.3cm of b2] (f1b) {\(\pi^+(p_3)\)};
    \vertex [below right=1.3cm of b2] (c2);
    \vertex [above right=1.2cm of c2] (f2b) {\(\pi^+(p_1)\)};
    \vertex [below right=1.2cm of c2] (f3b) {\(\pi^-(p_2)\)};

    \diagram* {
      (a2) -- [plain, line width=1pt] (b2),
      (b2) -- [plain, line width=1pt] (f2b),
      (b2) -- [plain, line width=1pt,edge label'=\(P-p_1\)] (c2), 
      (c2) -- [plain, line width=1pt] (f1b),
      (c2) -- [plain, line width=1pt] (f3b),
    };
      \end{feynman}
\end{tikzpicture}
\caption{\footnotesize{$D_s$ decay into three pions via scalar mesons. \label{f4}}}
\end{figure}
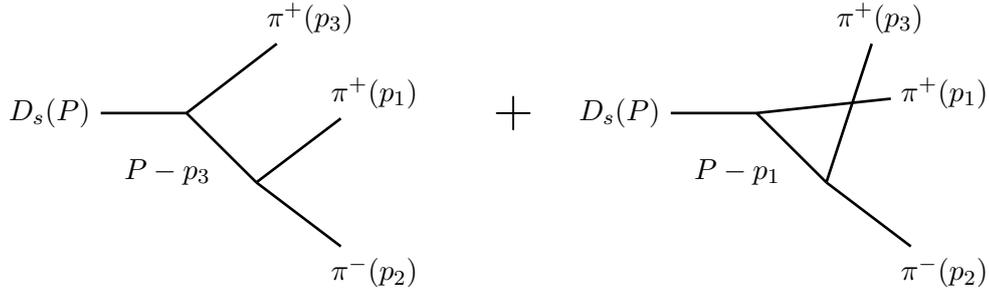

Differently from the case $D_s\to K\bar  K\pi$ (see Fig.~\ref{f3}), in $D_s\to 3\pi$ there are two interfering amplitudes (see  Fig.~\ref{f4})

The amplitude $M$ in~\eqref{emm2} has to be substituted by the sum of the two unstable particle propagators
\be
M=\frac{1}{m_f^2-m_{12}^2-i \gamma}+\frac{1}{ m_f^2-m_{23}^2-i \gamma}
\ee 
where again $\gamma=m_f\Gamma_f$. This is included in~\eqref{3body} to determine the rates of interest, with the integration measure in~\eqref{3body} divided by 2, for the identical particles.

First of all we can determine $A_I$ by requiring that $\Gamma(D_s\to \sigma\pi \to 3\pi):\Gamma(D_s\to f_0\pi\to 3\pi)\lesssim 0.1$ as reported by LHCb~\cite{LHCb:2022pjv}. The rate in the denominator is determined by the $B_f$ amplitude determined in the previous section, 
and we obtain
\be
|A_I|\leq 15.4 \label{lastamp}
\ee

For the other channels of $D_s\to 3\pi$, we predict
\begin{align}
& \Gamma(D_s\to f^\prime\pi \to 3\pi):\Gamma(D_s\to f_0\pi\to  3\pi)\simeq 0.8\\
&\Gamma(D_s\to f^{\prime\prime}\pi \to 3\pi):\Gamma(D_s\to f_0\pi\to 3\pi)\simeq 0.01
\end{align}
\begin{table}[ht]
\centering
    \begin{tabular}{|c|c|c|}
     \hline
Channel & {\footnotesize{Predicted Rates/Rate($f_0(980)$)}} & {\footnotesize{Expt.~\cite{ParticleDataGroup:2024cfk}}} \\ 
 \hline
 {\footnotesize{$\pi^+ \sigma(500)$}} &  {\footnotesize{ $<0.1$~\footnotesize{assumed for $A_I$}}}& {\footnotesize{not reported}}\\  
  \hline
 {\footnotesize{$\pi^+f_0(980)$}} & {\footnotesize{ $1$}}& {\footnotesize{$0.565\pm0.043\pm 0.047$}}    \\  
 \hline
{\footnotesize{$\pi^+ f_0(1370)$}} &  {\footnotesize{ $0.8$}}&{\footnotesize{$0.324\pm0.077\pm 0.017$}} \\ 
  \hline
{\footnotesize{$\pi^+ f_0(1710)$}} &  {\footnotesize{ $1\cdot 10^{-2} $}} &{\footnotesize{not reported}}\\  
\hline \hline
  \end{tabular}
  \caption{\footnotesize {Predicted rates, normalized to the rate of $f_0(980) \to \pi^+\pi^+\pi^-$and observed branching fractions.}}
  \label{FFpred1}
\end{table}

The presence of a $ u  \bar u+ d \bar d$ meson such as $f^\prime$ in the $D_s^+$, Cabibbo allowed, decays with final states dominated by $ s\bar s$ quark pairs, is quite remarkable. 

In our scheme, $f^\prime$ production arises quite naturally from the diagrams of Fig.~\ref{f2} (b) and (c) through its instanton generated four-quark component, see Eqs.~\eqref{mix} and~\eqref{instmix2}.
In the molecular model, it would require a sizeable violation of the Zweig rule, similar  to the violation required to explain the large $f_0\to \pi \pi$ decay of the $K\bar K$ molecule. 

\section{Cabibbo suppressed $S$-wave decay: $D^+\to \pi^+\pi^+\pi^-$\label{C_suppressedPi}}
\begin{figure}[h]
\centering
\scalebox{0.8}{
\begin{tikzpicture}
  \shade[ball color=gray!50] (0,0) circle (0.3);
  \node at (-0.7,0) {$D^+$};
   \node at (2,-1.7) {$(a)$};

  \begin{feynman}
    \vertex (a) at (0,0);
    \vertex (f1) at (3,1.2);
    \vertex (f2) at (3,-1.2);
    \vertex (f5) at (5,2.8);
    \vertex (f6) at (5,1.6);
    
     \node at ($(f1)+(0.3,0)$) {$d$};
      \node at ($(f2)+(0.3,0)$) {$\bar d$};
      
       \node at ($(f5)+(0.3,0)$) {$u$};
      \node at ($(f6)+(0.3,0)$) {$\bar d$};

    \vertex (mid) at ($(1,1.2)!0.5!(f1)$);
    \vertex (p) at ($(mid)+(1.5,1)$);

  \node at ($(a)+(3,0)$) {$J^P=0^+,\; L=1$};
        \node at ($(p)+(2.5,0)$) {$\pi^+$};

    \diagram* {
      (a) -- [fermion, out=90, in=180, thick] (f1),
      (a) -- [anti fermion, out=-90, in=180, thick ] (f2),
    
      (mid) -- [photon, thick] (p),
      
      (p) -- [fermion, out=90, in=180, thick] (f5),
      (p) -- [anti fermion, out=-90, in=180, thick ] (f6),        
    };
  \end{feynman}
\end{tikzpicture}
}
\caption{\footnotesize{$D$ decay, Cabibbo forbidden.\label{f5}}}
\end{figure}
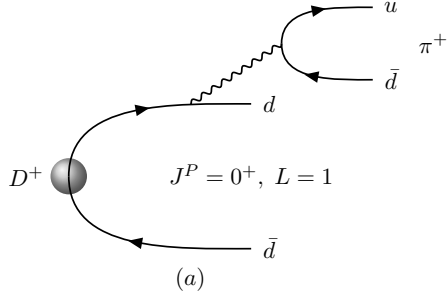

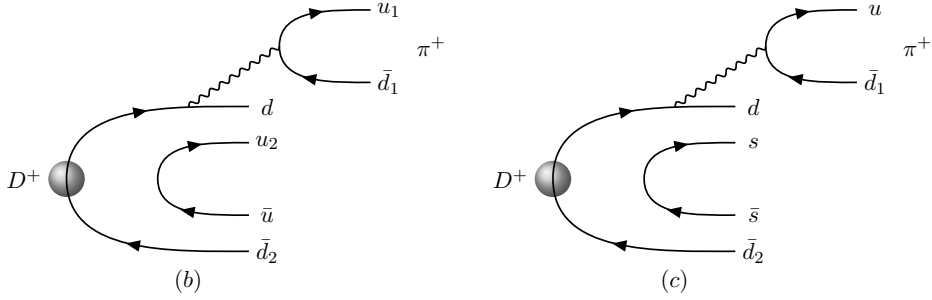
\begin{figure}[ht!]
\centering
\scalebox{0.8}{
\begin{tikzpicture}
  \begin{scope}[shift={(-4,0)}]
    \shade[ball color=gray!50] (0,0) circle (0.3);
    \node at (-0.7,0) {$D^+$};
       \node at (2,-1.7) {$(b)$};

    \begin{feynman}
      \vertex (a) at (0,0);
      \vertex (f1) at (3,1.2);
      \vertex (f2) at (3,-1.2);

      \vertex (b) at (1.5,0);
      \vertex (f3) at (3,0.6);
      \vertex (f4) at (3,-0.6);

      \vertex (mid) at ($(1,1.2)!0.5!(f1)$);
      \vertex (p) at ($(mid)+(1.5,1)$);
      \vertex (f5) at (5,2.8);
      \vertex (f6) at (5,1.6);

      \node at ($(f1)+(0.3,0)$) {$d$};
      \node at ($(f2)+(0.3,0)$) {$\bar d_2$};
      
      \node at ($(f3)+(0.3,0)$) {$u_2$};
      \node at ($(f4)+(0.3,0)$) {$\bar u$};
     
      \node at ($(f5)+(0.3,0)$) {$u_1$};
      \node at ($(f6)+(0.3,0)$) {$\bar d_1$};

   \node at ($(p)+(2.5,0)$) {$\pi^+$};

      \diagram* {
        (a) -- [fermion, out=90, in=180, thick] (f1),
        (a) -- [anti fermion, out=-90, in=180, thick] (f2),

        (b) -- [fermion, out=90, in=180, thick] (f3),
        (b) -- [anti fermion, out=-90, in=180, thick] (f4),

        (mid) -- [photon, thick] (p),
        (p) -- [fermion, out=90, in=180, thick] (f5),
        (p) -- [anti fermion, out=-90, in=180, thick] (f6),
      };
    \end{feynman}
  \end{scope}


  \begin{scope}[shift={(4,0)}]
    \shade[ball color=gray!50] (0,0) circle (0.3);
    \node at (-0.7,0) {$D^+$};
     \node at (2,-1.7) {$(c)$};

    \begin{feynman}
      \vertex (a) at (0,0);
      \vertex (f1) at (3,1.2);
      \vertex (f2) at (3,-1.2);

      \vertex (b) at (1.5,0);
      \vertex (f3) at (3,0.6);
      \vertex (f4) at (3,-0.6);

      \vertex (mid) at ($(1,1.2)!0.5!(f1)$);
      \vertex (p) at ($(mid)+(1.5,1)$);
      \vertex (f5) at (5,2.8);
      \vertex (f6) at (5,1.6);
      
       \node at ($(f1)+(0.3,0)$) {$d$};
      \node at ($(f2)+(0.3,0)$) {$\bar d_2$};
      
      \node at ($(f3)+(0.3,0)$) {$s$};
      \node at ($(f4)+(0.3,0)$) {$\bar s$};
     
      \node at ($(f5)+(0.3,0)$) {$u$};
      \node at ($(f6)+(0.3,0)$) {$\bar d_1$};
 	  \node at ($(p)+(2.5,0)$) {$\pi^+$};

      \diagram* {
        (a) -- [fermion, out=90, in=180, thick] (f1),
        (a) -- [anti fermion, out=-90, in=180, thick] (f2),

        (b) -- [fermion, out=90, in=180, thick] (f3),
        (b) -- [anti fermion, out=-90, in=180, thick] (f4),

        (mid) -- [photon, thick] (p),
        (p) -- [fermion, out=90, in=180, thick] (f5),
        (p) -- [anti fermion, out=-90, in=180, thick] (f6),
      };
    \end{feynman}
  \end{scope}
\end{tikzpicture}
}
\caption{\footnotesize{$D$ decay, Cabibbo forbidden, with a quark pair from the vacuum. \label{f6}}}
\end{figure}

We have a different set of quark diagrams for this case. They are reported in Fig.~\ref{f5} and Fig.~\ref{f6}. 
The basic features of these diagrams are the following
 \begin{enumerate}
 \item In $(a)$, $f^\prime$ is produced  (amplitude $A_f$) whereas $f_0$ is produced via mixing (amplitude $A_I$) 
 \item In $(b)$, $\sigma$ is produced (amplitude $B_f$) and, via mixing, $f^{\prime\prime}$ (amplitude $B_I$)  
 \item  The amplitude in $(c)$ is shared between  $f_0$  (amplitude $B_f$) and $a_0$, which, however, does not decay in $\pi^+\pi^-$
 \item The diagram in $(c)$ gives rise to the most important contribution to $f^{\prime}$ (amplitude $B_I$)
\end{enumerate}

We will now use the effective couplings
\bea
&& g_0=B_f \sqrt{m_D}\quad \text{for}~\sigma\notag \\
&& g_0=(B_f/\sqrt{2}~\pm~A_I/\sqrt{2})\sqrt{m_{D}}\quad \text{for}~f_0 \label{linamp0} \\
&& g_0=(B_I/\sqrt{2}~\pm~A_f/\sqrt{2})\sqrt{m_{D}}\quad \text{for}~f^\prime  \label{linamp} \\
&&g_0=B_I\sqrt{m_{D}}\quad\text{for}~f^{\prime\prime}\notag
\eea
where the amplitudes $A,B$ have been determined in the previous section. The following ratios are determined by taking the four sign alternatives in~\eqref{linamp0} and \eqref{linamp}. 
{\it Taking the $-$ sign}  in both~\eqref{linamp0} and~\eqref{linamp} we obtain
\begin{align}
& \Gamma(D\to f_0\pi \to 3\pi):\Gamma(D\to \sigma\pi\to  3\pi)\simeq 0.17
\\
&\Gamma(D\to f^{\prime}\pi \to 3\pi):\Gamma(D\to \sigma\pi\to 3\pi)\simeq 0.07
\\
& \Gamma(D\to f^{\prime\prime}\pi \to 3\pi):\Gamma(D\to \sigma\pi\to  3\pi)\simeq 0.01
\end{align}
which happens to match very well the available experimental determination based on CLEO~\cite{CLEO:2007ktn} data, as detailed in the following Table\footnote{The $(+,+),(+,-),(-,+),(-,-)$ sign choices correspond to values $(0.54, 0.56),(0.54, 0.07),(0.17, 0.56),(0.17, 0.07),$ respectively.
} 
\begin{table}[ht]
\centering
    \begin{tabular}{|c|c|c|}
     \hline
\footnotesize{ Channel} & {\footnotesize{Predicted Rates /Rate($\sigma(500)$)}} & {\footnotesize{Expt.($~10^{-3}$)~\cite{ParticleDataGroup:2024cfk,CLEO:2007ktn}}}
\\ 
 \hline
  {\footnotesize{$\pi^+ \sigma(500)$}} &  {\footnotesize{ $1$}} & {\footnotesize{$1.38\pm 0.1$}} \\  
  \hline
 {\footnotesize{$\pi^+f_0(980)$}} &\footnotesize{0.17}& {\footnotesize{$0.157\pm 0.032$}}    \\  
 \hline
 {\footnotesize{$\pi^+ f_0(1370)$}} &\footnotesize{0.07} &{\footnotesize{$0.08\pm 0.04$}} \\ 
  \hline
{\footnotesize{$\pi^+ f_0(1710)$}} &\footnotesize{0.02} &{\footnotesize{not reported}}\\  
\hline \hline
  \end{tabular}
  \caption{\footnotesize { $D^+\to 3\pi$. Predicted rates, normalized to the rate of $\sigma(500) \to \pi^+\pi^+\pi^-$and observed branching fractions. }}
  \label{FFpred2}
\end{table}

Recent Dalitz plots from the LHCb experiment~\cite{LHCb:2022lja} suggest that the signals for $\sigma$ and $f_0$ may have comparable intensities. This observation could potentially lead to a preference for a different sign choice in the equations~\eqref{linamp0} and~\eqref{linamp}. However, as of now, there is no LHCb determination available to replace the CLEO values in the Table.

\section{Cabibbo suppressed $S$-wave decays $D_s^+\to K^+\pi^+\pi^-$\label{C_suppressedK}}
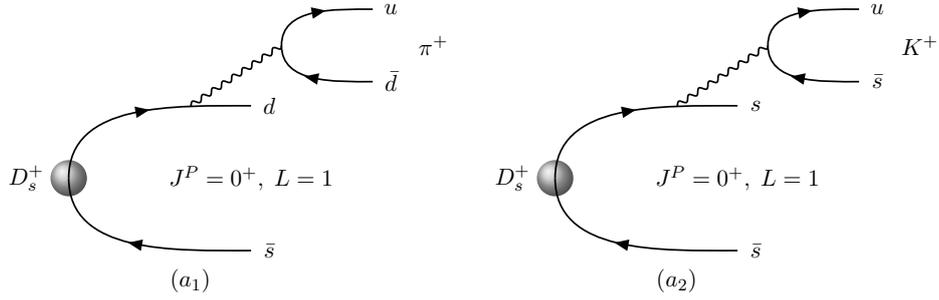
\begin{figure}[ht!]
\centering
\scalebox{0.8}{
\begin{tikzpicture}
  \begin{scope}[shift={(-4,0)}]
    \shade[ball color=gray!50] (0,0) circle (0.3);
    \node at (-0.7,0) {$D_s^+$};
  \node at (2,-1.7) {$(a_1)$};
  
    \begin{feynman}
      \vertex (a) at (0,0);
      \vertex (f1) at (3,1.2);
      \vertex (f2) at (3,-1.2);

      \vertex (b) at (1.5,0);
      \vertex (f3) at (3,0.6);
      \vertex (f4) at (3,-0.6);

      \vertex (mid) at ($(1,1.2)!0.5!(f1)$);
      \vertex (p) at ($(mid)+(1.5,1)$);
      \vertex (f5) at (5,2.8);
      \vertex (f6) at (5,1.6);

      \node at ($(f1)+(0.3,0)$) {$d$};
      \node at ($(f2)+(0.3,0)$) {$\bar s$};
      
     
      \node at ($(f5)+(0.3,0)$) {$u$};
      \node at ($(f6)+(0.3,0)$) {$\bar d$};

      \node at ($(p)+(2.5,0)$) {$\pi^+$};
 
  \node at ($(a)+(3,0)$) {$J^P=0^+,\; L=1$};

      \diagram* {
        (a) -- [fermion, out=90, in=180, thick] (f1),
        (a) -- [anti fermion, out=-90, in=180, thick] (f2),


        (mid) -- [photon, thick] (p),
        (p) -- [fermion, out=90, in=180, thick] (f5),
        (p) -- [anti fermion, out=-90, in=180, thick] (f6),
      };
    \end{feynman}
  \end{scope}


  \begin{scope}[shift={(4,0)}]
    \shade[ball color=gray!50] (0,0) circle (0.3);
    \node at (-0.7,0) {$D_s^+$};
    \node at (2,-1.7) {$(a_2)$};

    \begin{feynman}
      \vertex (a) at (0,0);
      \vertex (f1) at (3,1.2);
      \vertex (f2) at (3,-1.2);

      \vertex (b) at (1.5,0);
      \vertex (f3) at (3,0.6);
      \vertex (f4) at (3,-0.6);

      \vertex (mid) at ($(1,1.2)!0.5!(f1)$);
      \vertex (p) at ($(mid)+(1.5,1)$);
      \vertex (f5) at (5,2.8);
      \vertex (f6) at (5,1.6);
      
       \node at ($(f1)+(0.3,0)$) {$s$};
      \node at ($(f2)+(0.3,0)$) {$\bar s$};
      
     
      \node at ($(f5)+(0.3,0)$) {$u$};
      \node at ($(f6)+(0.3,0)$) {$\bar s$};

      \node at ($(p)+(2.5,0)$) {$K^+$};
      
     \node at ($(a)+(3,0)$) {$J^P=0^+,\; L=1$};

      \diagram* {
        (a) -- [fermion, out=90, in=180, thick] (f1),
        (a) -- [anti fermion, out=-90, in=180, thick] (f2),


        (mid) -- [photon, thick] (p),
        (p) -- [fermion, out=90, in=180, thick] (f5),
        (p) -- [anti fermion, out=-90, in=180, thick] (f6),
      };
    \end{feynman}
  \end{scope}
\end{tikzpicture}
}

\caption{$D_s$ Cabibbo suppressed quark diagrams.\label{f7}}
\end{figure}

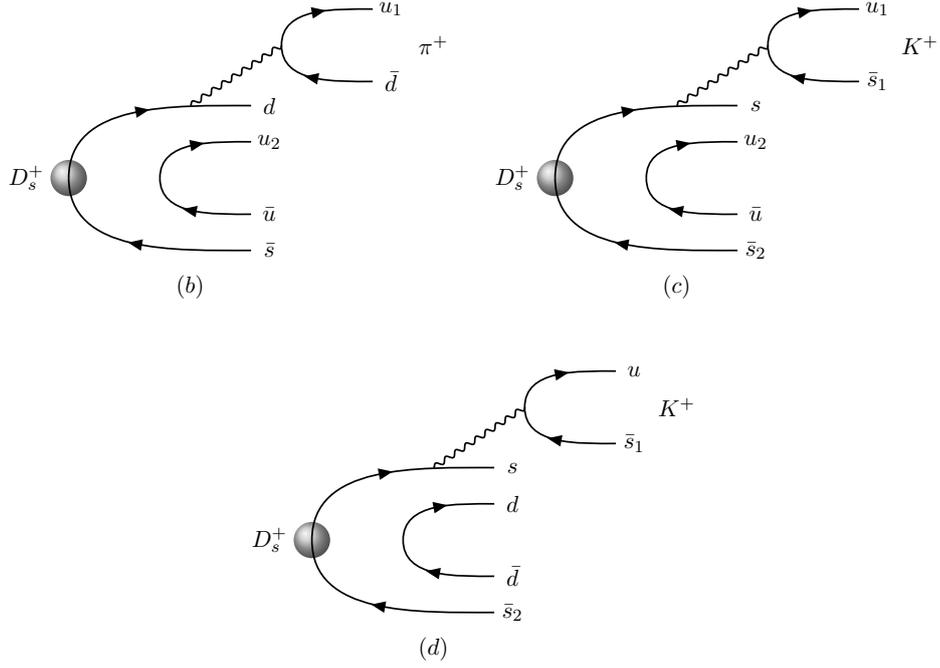
\begin{figure}[ht!]
\centering
\scalebox{0.8}{
\begin{tikzpicture}
  \begin{scope}[shift={(-4,0)}]
    \shade[ball color=gray!50] (0,0) circle (0.3);
    \node at (-0.7,0) {$D_s^+$};

    \begin{feynman}
      \vertex (a) at (0,0);
      \vertex (f1) at (3,1.2);
      \vertex (f2) at (3,-1.2);

      \vertex (b) at (1.5,0);
      \vertex (f3) at (3,0.6);
      \vertex (f4) at (3,-0.6);

      \vertex (mid) at ($(1,1.2)!0.5!(f1)$);
      \vertex (p) at ($(mid)+(1.5,1)$);
      \vertex (f5) at (5,2.8);
      \vertex (f6) at (5,1.6);

      \node at ($(f1)+(0.3,0)$) {$d$};
      \node at ($(f2)+(0.3,0)$) {$\bar s$};
      
      \node at ($(f3)+(0.3,0)$) {$u_2$};
      \node at ($(f4)+(0.3,0)$) {$\bar u$};
     
      \node at ($(f5)+(0.3,0)$) {$u_1$};
      \node at ($(f6)+(0.3,0)$) {$\bar d$};

      \node at ($(p)+(2.5,0)$) {$\pi^+$};

      \node at ($(mid)+(0,-3)$) {$(b)$};

      \diagram* {
        (a) -- [fermion, out=90, in=180, thick] (f1),
        (a) -- [anti fermion, out=-90, in=180, thick] (f2),

        (b) -- [fermion, out=90, in=180, thick] (f3),
        (b) -- [anti fermion, out=-90, in=180, thick] (f4),

        (mid) -- [photon, thick] (p),
        (p) -- [fermion, out=90, in=180, thick] (f5),
        (p) -- [anti fermion, out=-90, in=180, thick] (f6),
      };
    \end{feynman}
  \end{scope}

  \begin{scope}[shift={(0,-6)}]
    \shade[ball color=gray!50] (0,0) circle (0.3);
    \node at (-0.7,0) {$D_s^+$};

    \begin{feynman}
      \vertex (a) at (0,0);
      \vertex (f1) at (3,1.2);
      \vertex (f2) at (3,-1.2);

      \vertex (b) at (1.5,0);
      \vertex (f3) at (3,0.6);
      \vertex (f4) at (3,-0.6);

      \vertex (mid) at ($(1,1.2)!0.5!(f1)$);
      \vertex (p) at ($(mid)+(1.5,1)$);
      \vertex (f5) at (5,2.8);
      \vertex (f6) at (5,1.6);

      \node at ($(f1)+(0.3,0)$) {$s$};
      \node at ($(f2)+(0.3,0)$) {$\bar s_2$};
      
      \node at ($(f3)+(0.3,0)$) {$d$};
      \node at ($(f4)+(0.3,0)$) {$\bar d$};
     
      \node at ($(f5)+(0.3,0)$) {$u$};
      \node at ($(f6)+(0.3,0)$) {$\bar s_1$};

      \node at ($(p)+(2.5,0)$) {$K^+$};
 
   \node at ($(mid)+(0,-3)$) {$(d)$};

      \diagram* {
        (a) -- [fermion, out=90, in=180, thick] (f1),
        (a) -- [anti fermion, out=-90, in=180, thick] (f2),

        (b) -- [fermion, out=90, in=180, thick] (f3),
        (b) -- [anti fermion, out=-90, in=180, thick] (f4),

        (mid) -- [photon, thick] (p),
        (p) -- [fermion, out=90, in=180, thick] (f5),
        (p) -- [anti fermion, out=-90, in=180, thick] (f6),
      };
    \end{feynman}
  \end{scope}


  \begin{scope}[shift={(4,0)}]
    \shade[ball color=gray!50] (0,0) circle (0.3);
    \node at (-0.7,0) {$D_s^+$};

    \begin{feynman}
      \vertex (a) at (0,0);
      \vertex (f1) at (3,1.2);
      \vertex (f2) at (3,-1.2);

      \vertex (b) at (1.5,0);
      \vertex (f3) at (3,0.6);
      \vertex (f4) at (3,-0.6);

      \vertex (mid) at ($(1,1.2)!0.5!(f1)$);
      \vertex (p) at ($(mid)+(1.5,1)$);
      \vertex (f5) at (5,2.8);
      \vertex (f6) at (5,1.6);
      
       \node at ($(f1)+(0.3,0)$) {$s$};
      \node at ($(f2)+(0.3,0)$) {$\bar s_2$};
      
      \node at ($(f3)+(0.3,0)$) {$u_2$};
      \node at ($(f4)+(0.3,0)$) {$\bar u $};
     
      \node at ($(f5)+(0.3,0)$) {$u_1$};
      \node at ($(f6)+(0.3,0)$) {$\bar s_1$};

      \node at ($(p)+(2.5,0)$) {$K^+$};
      
        \node at ($(mid)+(0,-3)$) {$(c)$};

      \diagram* {
        (a) -- [fermion, out=90, in=180, thick] (f1),
        (a) -- [anti fermion, out=-90, in=180, thick] (f2),

        (b) -- [fermion, out=90, in=180, thick] (f3),
        (b) -- [anti fermion, out=-90, in=180, thick] (f4),

        (mid) -- [photon, thick] (p),
        (p) -- [fermion, out=90, in=180, thick] (f5),
        (p) -- [anti fermion, out=-90, in=180, thick] (f6),
      };
    \end{feynman}
  \end{scope}
\end{tikzpicture}
}
\caption{\footnotesize{$D_s$ Cabibbo suppressed quark diagrams with pair production from the vacuum. \label{f8} }}
\end{figure}

The relevant quark diagrams are reported in Fig.~\ref{f7} and~\ref{f8}. These are the main features to take into account 
\begin{enumerate}
 \item $(a_1)$ produces $K_0(1430)$ via $A_f$, and  $\kappa$ via $A_I$
  \item $(a_2)$ produces $f^{\prime\prime}$  via $A_f$ and $\sigma$ via  $A_I$
 \item $(b)$ produces $\kappa$ via $B_f$ and $K_0(1430)$ via $B_I$ 
 \item $(c, d)$ produce $f_0$ via  $\sqrt{2}B_f$ and $f^\prime$ via $\sqrt{2}B_I$ | compare to~\eqref{44} and~\eqref{45}
\end{enumerate}

Using the same methods as in the previous Sections  we find
\begin{align}
& \Gamma(D\to \sigma K \to \pi\pi K):\Gamma_0\simeq 0.05 \label{uu}\\
&\Gamma(D\to \kappa \pi \to \pi\pi K):\Gamma_0\simeq 0.28\\
& \Gamma(D\to f^{\prime}K \to \pi\pi K):\Gamma_0\simeq 0.43 \\
& \Gamma(D\to f^{\prime\prime}K \to \pi\pi K):\Gamma_0\simeq 0.001\\
& \Gamma(D\to K_0(1430)\pi \to \pi\pi K):\Gamma_0\simeq 0.31 \label{cc}
\end{align}
where we have used,  in the order
 \be 
 g_0=A_I\sqrt{m_{D_s}}, ~( B_f - A_I)
 \sqrt{m_{D_s}},~B_I\sqrt{2m_{D_s}},~A_f\sqrt{m_{D_s}}, ~( B_I - A_f)\sqrt{m_{D_s}} \label{sc}
 \ee
Ratios are taken with respect to $\Gamma_0$
\be
\Gamma_0=\Gamma(D\to f_0 K\to  \pi\pi K)\
\ee
where $g_0=B_f\sqrt{2m_{D_s}}$. 
These results are compared to experimental data from \cite{BESIII:2022vaf} in Table~\ref{FFKpp}, employing the same sign convention (see Eq.~\eqref{sc}) as utilized in the preceding section. No alternative sign convention facilitates a more favorable comparison to the data. We  suspect that a significant portion of this discrepancy  may be attributable to the fact that the $\kappa$ is not even incorporated into the experimental fits.  

\begin{table}[ht]
\centering
    \begin{tabular}{|c|c|c|}
     \hline
    
 \footnotesize{Channel} & \footnotesize{ Predicted Rates/Rate($f_0(980)$)}& {\footnotesize{ Expt. BR($10^{-4}$)~ \cite{ParticleDataGroup:2024cfk,BESIII:2022vaf}}} \\  \hline
  {\footnotesize{$K^+ \sigma(500)$}} &  {\footnotesize{ $0.05$}} &{\footnotesize{ $4.5\pm 3.0$}}\\  \hline
{\footnotesize{$K^+f_0(980)$}} & {\footnotesize{ $1$}} & {\footnotesize{$2.8\pm 1.1$}}\\  \hline
{\footnotesize{$K^+ f_0(1370)$}} &  {\footnotesize{ $0.43$}} &{\footnotesize{ $12 \pm 6$}}\\  \hline
 {\footnotesize{$K^+ f_0(1710)$}} &  {\footnotesize{ $0.001$}} &{\footnotesize{not reported}}\\  \hline
{\footnotesize{$\pi^+ \kappa(700)$}} & {\footnotesize{0.28}} &{\footnotesize{ not considered in fit}}\\  \hline
{\footnotesize{$\pi^+ K_0(1430)$}} &{\footnotesize{ $0.31$}} & {\footnotesize{ $9.4 \pm 3.2$}}\\  \hline
 \hline
  \end{tabular}
  \caption{\footnotesize {Predicted rates for  $D^+\to K^+\pi^+\pi^-$, normalised to the $f_0$ rate.
The comparison of our predictions  with data from\cite{ParticleDataGroup:2024cfk,BESIII:2022vaf} is not good despite the rather large experimental errors. Also, unlike previous cases, BESIII fits the low energy $K^+\pi^-$ data by a continuum distribution, rather then the Breit-Wigner corresponding to the $\kappa$ resonance, as would be required by our model.}}
  \label{FFKpp}
  
\end{table}

\section{Conclusions} \label{concl}
In~\cite{tHooft:2008rus}, it is proposed that 6-fermion instanton-induced interactions can significantly contribute to the dynamics of scalar mesons at low energies. We further investigate this possibility within the context of recently obtained LHCb experimental data on $D$ and $D_s$ decays into three pseudoscalars. We substantially corroborate this picture, yet we encounter difficulties with the Cabibbo suppressed $S$-wave decays 
$D_s^+\to K^+\pi^+\pi^-$ that may be attributed to the treatment of the $\kappa(700)$ resonance in data analyses or to characteristics of the model beyond our control.
We believe that the present study represents a  step in favor of a tetraquark interpretation of the lowest scalar meson nonet.

\section*{Acknowledgements \label{acknow}}
We acknowledge  illuminating and very informative discussions with V. Belyaev, G. Cavoto, D. Germani, Feng-Kun Guo, Gino Isidori and  Xiaoyan Shen.  L.M. and V. R. express their gratitude to G. Giudice and members of the CERN TH-Division, where a significant portion of the present study was done, for insightful conversations.

\appendix

\section*{Appendix: Rules for amplitudes  \label{rules}}
We summarize the rules for constructing the amplitudes  of different processes, from the basic amplitudes $A_f,~A_I,~B_f,~B_I$ defined in Sects.~\ref{C_allowedK} and \ref{C_allowedPi}.
\\\\
Diagrams in Sect.~\ref{C_allowedK}:
\begin{enumerate}
\item Diagram Fig.~\ref{f1} (a): An $f^{\prime\prime}$  is produced from an  $s\bar s$ pair through the  $A_f$ amplitude 
\item Diagram Fig.~\ref{f2} (b): The quark combination $[su][ \bar s\bar u]$ produces a $f_0=([su][ \bar s\bar u]+[sd][ \bar s\bar d])/\sqrt{2}$   through the amplitude  $B_f/\sqrt{2}$. If the same quark combination is supposed to produce  $f^\prime$ it will go through the instanton amplitude $B_I/\sqrt{2}$.  Same for Diagram Fig.~\ref{f1} (c).
\end{enumerate}
Diagrams in  Sect.~\ref{C_allowedPi}:
Same quark diagrams of Sect.~\ref{C_allowedK}, same amplitudes to produce $f^{\prime\prime}$ and $f_0,~f^\prime$ but now also $\sigma$ is produced in Fig.~\ref{f1} (a) with amplitude $A_I$.
\\\\
Diagrams in  Sect.\ref{C_suppressedPi}
\begin{enumerate}
\item Diagram Fig.~\ref{f5}~(a):  Gives $  f^{\prime}$ through $A_f/\sqrt{2}$ and $ f_0$ through $A_I/\sqrt{2}$
\item Diagram of Fig. \ref{f6}~(b): The quark combination $[ud][\bar u \bar d]$ produces the $\sigma$ through the  $B_f$ amplitude and $f^{\prime\prime}$ through the amplitude $B_I$
\item  Diagram of Fig. \ref{f6}~(c): Gives $f_0$ through the amplitude  $B_f/\sqrt{2}$ and $f^\prime$ through the amplitude  $B_I/\sqrt{2}$
\end {enumerate}
\leavevmode\\
Diagrams in  Sect.~\ref{C_suppressedK}:
\begin{enumerate}
\item Diagram Fig.~\ref{f7} ($a_1$):  Gives $K_0(1430)$ via $d\bar s $ through $A_f$ and $\kappa$ via $A_I$
\item Diagram Fig.~\ref{f7} ($a_2$):  Gives $f^{\prime\prime}$ via $A_f$ and $\sigma$ via $A_I$
\item Diagram Fig.~\ref{f8} ($b$): From $[ud] [\bar u \bar s]$ gives $\kappa$  via $B_f$ and $K_0(1430)$ via $B_I$
\item Diagram Fig.~\ref{f8}: ($c$) $f_0/\sqrt{2}$   through the amplitude  $B_f/\sqrt{2}$ and $f^\prime/\sqrt{2}$ through $B_I/\sqrt{2}$. Same for diagram (d)
\end{enumerate}

 \end{document}